![JOSS - The Journal of Open Source Software]# mcdust: A 2D Monte Carlo code for dust coagulation in protoplanetary disks

Vignesh Vaikundaraman [1], Nerea Gurrutxaga [1], and Joanna Drążkowska [1]

**1** Max Planck Institute for Solar System Research, Justus-von-Liebig-Weg 3 37077 Gottingen
**DOI:** 10.xxxxx/draft

**Software**
- Review ↗
- Repository ↗
- Archive ↗

**Editor:** ↗
**Submitted:** 11 July 2025
**Published:** unpublished

**License**
Authors of papers retain copyright and release the work under a Creative Commons Attribution 4.0 International License (CC BY 4.0).
## Summary

`mcdust` is a parallel simulation code for dust evolution in protoplanetary disks. The code is written in `FORTRAN90` and parallelized with `OpenMP`. The code models dust collisional evolution and transport in the vertical and radial directions. The currently included collisional outcomes are dust growth by sticking, fragmentation of dust particles and erosion, where a small particle chips a portion of the large particle. We employ a representative particle approach detailed in Zsom & Dullemond (2008) to track a limited number of particles instead of tracking every particle, saving computational time. We have a static power-law gas disk with temperature assumed to be vertically isothermal. Dust coagulation depends on the local gas properties and therefore we bin particles into grids and perform collisions. We make use of an adaptive grid approach where we make sure that each cell has equal number of representative particles. This guarantees that there are always sufficient particles to resolve the physics of collisions. Figure below shows a sketch of our adaptive grid model and the different physical processes simulated by `mcdust`. The details of the physics of the code is explained in Drążkowska et al. (2013).
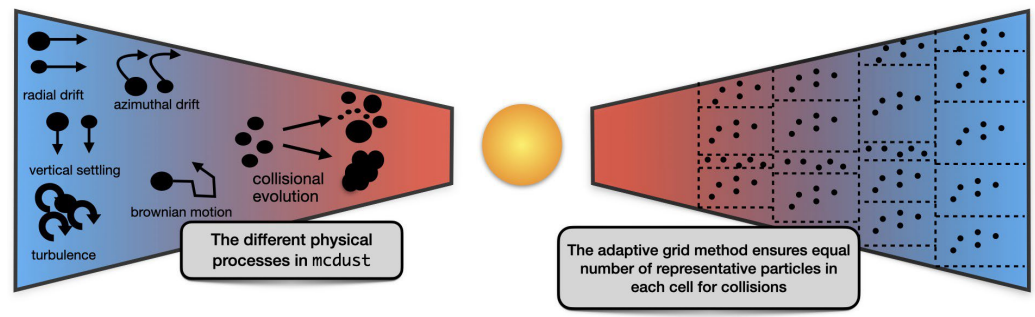

## Statement of Need

Modelling dust coagulation and evolution is an essential part of understanding planet formation and protoplanetary disks. The evolution of micrometer-sized dust grains to millimeter-sized grains sets the tone for planetesimal formation. Dust evolution is inextricably linked with the formation of substructures in disks. Dust dynamics is also of prime importance when understanding protoplanetary disk chemistry. Therefore very important to understand in detail the processes involved in dust growth and dynamics and how they influence different aspects of disk dynamics and planet growth.

Dust coagulation is generally modelled with the Smoluchowski equation, an integro-differential equation. Some examples of open source codes that solve the Smoluchowski equation are `dustpy` (Stammler & Birnstiel, 2022), a 1D code to simulate gas and dust evolution and `cuDisc`(Robinson et al., 2024), a 2D code modelling dust coagulation and disk evolution. Both

Vaikundaraman et al. (2025). mcdust: A 2D Monte Carlo code for dust coagulation in protoplanetary disks. *Journal of Open Source Software*, 1
¿VOL?(¿ISSUE?), 8619. https://doi.org/10.xxxxx/draft.

of these codes are Eulerian in nature, i.e. they follow volume rather than mass. The drawback with these methods is that it is difficult to track histories of dust particles. Furthermore, adding a property to track adds a dimension to solving the Smoluchowski Equation (Stammler et al., 2017). Lagrangian methods can solve this issue partly by allowing us to track histories of particles. `PHANTOM` (Price et al., 2018), is an open source Smoothed Particle Hydrodynamics (SPH) code that models models dust growth using a monodisperse growth or a 'single-size' approximation and do not solve the Smoluchowski equation for dust growth (Vericel et al., 2021).

Alternatively one can model dust coagulation by performing Monte Carlo simulations. Monte Carlo methods are well suited for stochastic processes and that makes it a suitable method for dust coagulation. The representative particle approach is Lagrangian in nature, meaning we track the particles and hence their histories. We can add properties to the particles that can be tracked without much computational complexity. This advantage is very useful when we want to combine dust evolution and chemistry, where we want to look at the histories of particles and their compositions when chemical models are included. We point the reader to Vaikundaraman et al. (2025) for an example of tracking composition/chemistry using `mcdust` to investigate the carbon depletion puzzle in the inner Solar System.

Including dust growth/dynamics in hydrodynamic simulations of protoplanetary disks can be very expensive and computationally complex. This is where `mcdust` can also be used to post process data from hydrodynamic simulations to understand the dynamics and evolution of dust in different conditions without much computational complexity.

Currently, there is no available open source code that models dust coagulation using Monte Carlo methods and such a code would be very helpful as an alternative way for modelling dust coagulation.

## Benchmark

We compare a run of our code with a run from the open source 1D dust coagulation code `dustpy`(Stammler & Birnstiel, 2022). The parameters used for the simulations are listed in the table below. The gas evolution in both the codes was switched off and the codes were run with a static gas background to exclude any differences in gas treatment that might influence the outcome of dust evolution.

| Parameter | Value |
| --- | --- |
| gas surface density at 1 AU | 1000 g/cm$^2$ |
| temperature at 1 AU | 280 K |
| turbulence strength ($\alpha$) | 10$^{-3}$ |
| fragmentation velocity ($v_{\mathbf{frag}}$) | 10 m/s |
| erosion mass ratio | 10 |
| gas surface density power law $p$ | 1 |
| temperature power law $q$ | 0.5 |

We ran a simulation for 10000 years. We show below the dust surface density $\sigma_d$ as a function of particle mass and distance from star at the end of the simulation for both `mcdust` and `dustpy`.

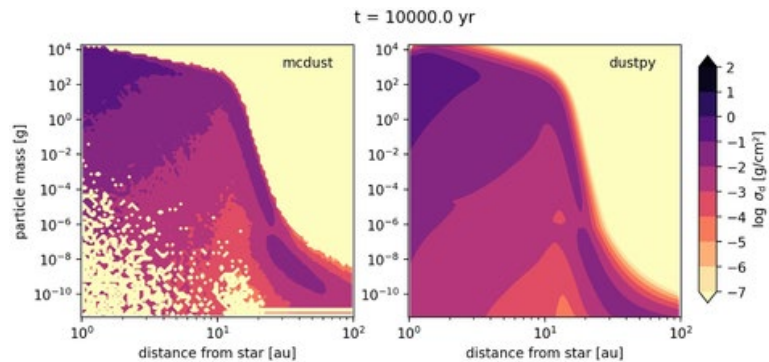

The image above shows the radial distrbution of It is evident that both the codes have similar overall outcomes but they do have certain differences. The most striking one is that `mcdust` does not provide coverage of the regions of parameter space that do not include sufficiently high fraction of the dust mass. This holds true for all Monte Carlo based codes. But with higher resolution simulation this issue can be overcome. The other important factor is that `mcdust` does not face the issue of artificially sped-up growth that `dustpy` and other Smoluchowski equation based codes tend to encounter. And the 2D r-z structure of `mcdust` also helps us to investigate processes like sedimentation driven coagulation (Drążkowska et al., 2013) that are not usually seen in 1D simulations like `dustpy`. This can be seen in image at around 50 AU where `mcdust` has larger surface densities higher masses when compared to `dustpy`. For a more detailed discussion of the differences between the two approaches to dust growth, i.e., the Monte Carlo method and the Smoluchowski equation approach we point the reader to Drążkowska et al. (2014).

# Acknowledgments

The authors acknowledge funding from the European Union under the European Union's Horizon Europe Research & Innovation Programme 101040037 (PLANETOIDS). Views and opinions expressed are however those of the author(s) only and do not necessarily reflect those of the European Union or the European Research Council. Neither the European Union nor the granting authority can be held responsible for them.